\documentclass{PoS}
 
\usepackage{epsfig,amsmath}

\title{Prospects of Jet Tomography Using Hard Processes inside a Soft Medium}

\ShortTitle{Prospects of Jet Tomography Using Hard Processes inside a Soft Medium}

\author{\speaker{Thorsten Renk}%
 
	\\
 
        Department of Physics, P.O. Box 35 FI-40014 University of Jyv\"askyl\"a, Finland and \\
	Helsinki Institute of Physics, P.O. Box 64 FI-00014, University of Helsinki, Finland\\
 
        E-mail: \email{trenk@phys.jyu.fi}}

\author{Kari J.~Eskola\\
 
       Department of Physics, P.O. Box 35 FI-40014 University of Jyv\"askyl\"a, Finland and \\
	Helsinki Institute of Physics, P.O. Box 64 FI-00014, University of Helsinki, Finland\\

        E-mail: \email{kari.eskola@phys.jyu.fi}}

\abstract{The term 'tomography' is commonly applied to the idea of studying properties of a medium by the modifications this medium induces to a known probe propagating through it. In the context of ultrarelativistic heavy-ion collisions, rare high transverse momentum ($p_T$) processes taking place alongside soft bulk-matter production can be viewed as a tomographic probe as long as the energy scales are such that the modification of high $p_T$ processes can be dominantly ascribed to interactions with the medium during the propagation of partons.
Various high $p_T$ observables have been suggested for tomography, among them hard single hadron suppression, dihadron correlations and $\gamma$-hadron correlations. In this paper, we present a model study of a number of different observables within the same calculational framework to assess the sensitivity of the observables to different properties of the medium and discuss the prospects of obtaining tomographic information.
}

\FullConference{High-$p_T$ physics at LHC\\
 
		 March 23-27, 2007 \\
 
		 University of Jyv\"{a}skyl\"{a}, Jyv\"{a}skyl\"{a}, Finland}

\begin{document}

\section{Introduction}
 
The expression 'jet tomography' is often used to describe the analysis of hard perturbative Quantum Chromodynamics (pQCD) processes taking place inside the soft matter created in an ultrarelativistic heavy-ion collision. Such processes, which are well understood in p-p collisions, can be viewed as a known probe as they take place before any formation scale of a soft medium. Thus, only the subsequent propagation of partons through the soft medium (and possibly hadronization, although at sufficiently high $p_T$ the hadron formation length is larger than the medium extension) is sensitive to interactions with the medium. Hence, any modification of hard processes embedded in a medium potentially carries tomographic information about the medium properties. 

In particular the experimental focus is on the nuclear suppression of hard hadrons in A-A collisions compared with the scaled baseline from p-p collisions, which is expected due to interactions of a hard parton with the soft medium (see e.g. \cite{Tomo}). 
However, the nuclear suppression factor 
\begin{equation}
R_{AA}(p_T,y) = \frac{d^2N^{AA}/dp_Tdy}{T_{AA}({\bf b}) d^2 \sigma^{NN}/dp_Tdy}.
\end{equation}
is a rather integral quantity, arising in model calculations from a convolution of the hard pQCD vacuum cross section $d\sigma_{vac}^{AA \rightarrow f +X}$ for the production of a  parton $f$, the energy loss probability $P_f(\Delta E)$ given the vertex position and path through the medium and the vacuum fragmentation function $D_{f \rightarrow h}^{vac}(z, \mu_F^2)$, as schematically expressed 
\begin{equation}
\label{E-Folding}
d\sigma_{med}^{AA\rightarrow h+X} = \sum_f d\sigma_{vac}^{AA \rightarrow f +X} \otimes P_f(\Delta E) \otimes
D_{f \rightarrow h}^{vac}(z, \mu_F^2),
\end{equation}
where
\begin{equation}
d\sigma_{vac}^{AA \rightarrow f +X} = \sum_{ijk} f_{i/A}(x_1,Q^2) \otimes f_{j/A}(x_2, Q^2) \otimes \hat{\sigma}_{ij 
\rightarrow f+k}.
\end{equation}
Here, $f_{i/A}(x, Q^2)$ denotes the nuclear parton distribution function which depends on the parton
momentum fraction $x$ and the hard momentum scale $Q^2$ and $\hat{\sigma}_{ij\rightarrow f+k}$ is the the partonic pQCD cross section.

Eq.~(\ref{E-Folding}) has to be properly averaged over all possible vertices distributed according to the nuclear overlap $T_{AA}$ and all possible paths through the medium. In \cite{gamma-h} we have argued that one can factorize this spatial averaging from the momentum space formulation Eq.~(\ref{E-Folding}) and thus define the geometry-averaged energy loss probability $\langle P(\Delta E, E) \rangle_{T_{AA}}$. $R_{AA}$ can thus be viewed as providing constraints for the form of  $\langle P(\Delta E, E) \rangle_{T_{AA}}$.

\section{Calculational framework}

Any model for medium modifications of a hard process must contain three major ingredients: The hard pQCD process, the bulk matter evolution for which we either use a hydrodynamic 
\cite{Hydro} or a parametrized evolution model \cite{Parametrized}  and the energy loss probability distribution 
given a hard parton path through the soft medium \cite{QuenchingWeights}. 

The primary hard process is calculated in leading order pQCD under the assumption that the transverse momentum scale is large enough so that hadronization takes place outside the medium and that the produced leading hadron can be assumed to be collinear with its parent parton. The calculation, when supplemented by a K-factor, agrees well with hard hadron production measured in p-p collisions. In particular, the AKK set of fragmentation functions \cite{AKK} also gives a satisfactory description of proton production whereas the older KKP set \cite{KKP} does not. Explicit expressions for the hard process calculation can be found e.g. in \cite{Correlations}.

The interaction of the hard parton with the soft medium is calculated using the radiative energy loss formalism of \cite{QuenchingWeights}. If we call the angle between outgoing parton and the reaction plane $\phi$, 
the path of a given parton through the medium $\xi(\tau)$ is specified 
by $({\bf r_0}, \phi)$ and we can compute the energy loss 
probability $P(\Delta E)_{path}$ for this path. We do this by 
evaluating the line integrals
\begin{equation}
\label{E-omega}
\omega_c({\bf r_0}, \phi) = \int_0^\infty \negthickspace d \xi \xi \hat{q}(\xi) \quad  \text{and} \quad \langle\hat{q}L\rangle ({\bf r_0}, \phi) = \int_0^\infty \negthickspace d \xi \hat{q}(\xi)
\end{equation}
along the path where we assume the relation
\begin{equation}
\label{E-qhat}
\hat{q}(\xi) = K \cdot 2 \cdot \epsilon^{3/4}(\xi) (\cosh \rho - \sinh \rho \cos\alpha)
\end{equation}
between the local transport coefficient $\hat{q}(\xi)$ (specifying 
the quenching power of the medium), the energy density $\epsilon$ and the local flow rapidity $\rho$ with angle $\alpha$ between flow and parton trajectory \cite{Flow1,Flow2}.
Here $\omega_c$ is the characteristic gluon frequency, setting the scale of the energy loss probability distribution, and $\langle \hat{q} L\rangle$ is a measure of the path-length weighted by the local quenching power.
We view 
the parameter $K$ as a tool to account for the uncertainty in the selection of $\alpha_s$ and possible non-perturbative effects increasing the quenching power of the medium (see discussion in \cite{Correlations}) and adjust it such that pionic $R_{AA}$ for central Au-Au collisions is described. Using the numerical results of \cite{QuenchingWeights}, we obtain $P(\Delta E; \omega_c, R)_{path}$ 
for $\omega_c$ and $R=2\omega_c^2/\langle\hat{q}L\rangle$ as a function of jet production vertex and the angle $\phi$.

The information about the soft medium is contained in the local energy density $\epsilon(\xi)$ and the flow rapidity $\rho(\xi)$. These parameters are obtained from dynamical evolution models which are tuned to describe a large body of bulk matter observables \cite{Hydro,Parametrized}. Details of the evolution models including contour plots of their time evolution can be found in \cite{Correlations}. In the following, we mainly illustrate three scenarios: A hydrodynamical evolution of matter ('Hydrodynamics'), the best fit to soft hadronic $p_T$ spectra and HBT correlation data of the parametrized evolution model ('Box density') and the hydrodynamical model under the assumption that only the partonic evolution phase leads to energy loss ('Black core'). Since in all models $R_{AA}$ for central Au-Au collisions is described by construction via a fit of $K$, the latter model implies that $K$ takes large values and the evolution exhibits a very black interior region and a dilute hadronic halo which does not induce energy loss at all, quite different from the other models.

\section{Single Hadron Suppression}

Since $R_{AA}$ does not contain any spatial information, the production vertices of hard partons and their path through the medium have to be averaged out. Hard vertices $(x_0,y_0)$ are distributed according to a probability density
\begin{equation}
\label{E-PGeo}
P(x_0,y_0) = \frac{T_{A}({\bf r_0 + b/2}) T_A(\bf r_0 - b/2)}{T_{AA}({\bf b})},
\end{equation}
where ${\bf b}$ is the impact parameter. The thickness function is given by the nuclear density
$\rho_A({\bf r},z)$ as $T_A({\bf r})=\int dz \rho_A({\bf r},z)$. Hence, given the energy loss probability distribution $P_f(\Delta E)_{path}$ for a given path through the medium, we obtain

\begin{equation}
\label{E-Pav}
\langle P_f(\Delta E, E)\rangle_{T_{AA}} \negthickspace =  \negthickspace \frac{1}{2\pi} \int_0^{2\pi} 
\negthickspace \negthickspace \negthickspace d\phi 
\int_{-\infty}^{\infty} \negthickspace \negthickspace \negthickspace \negthickspace dx_0 
\int_{-\infty}^{\infty} \negthickspace \negthickspace \negthickspace \negthickspace dy_0 P(x_0,y_0)  
P_f(\Delta E)_{path}.
\end{equation}

\begin{figure}
\epsfig{file=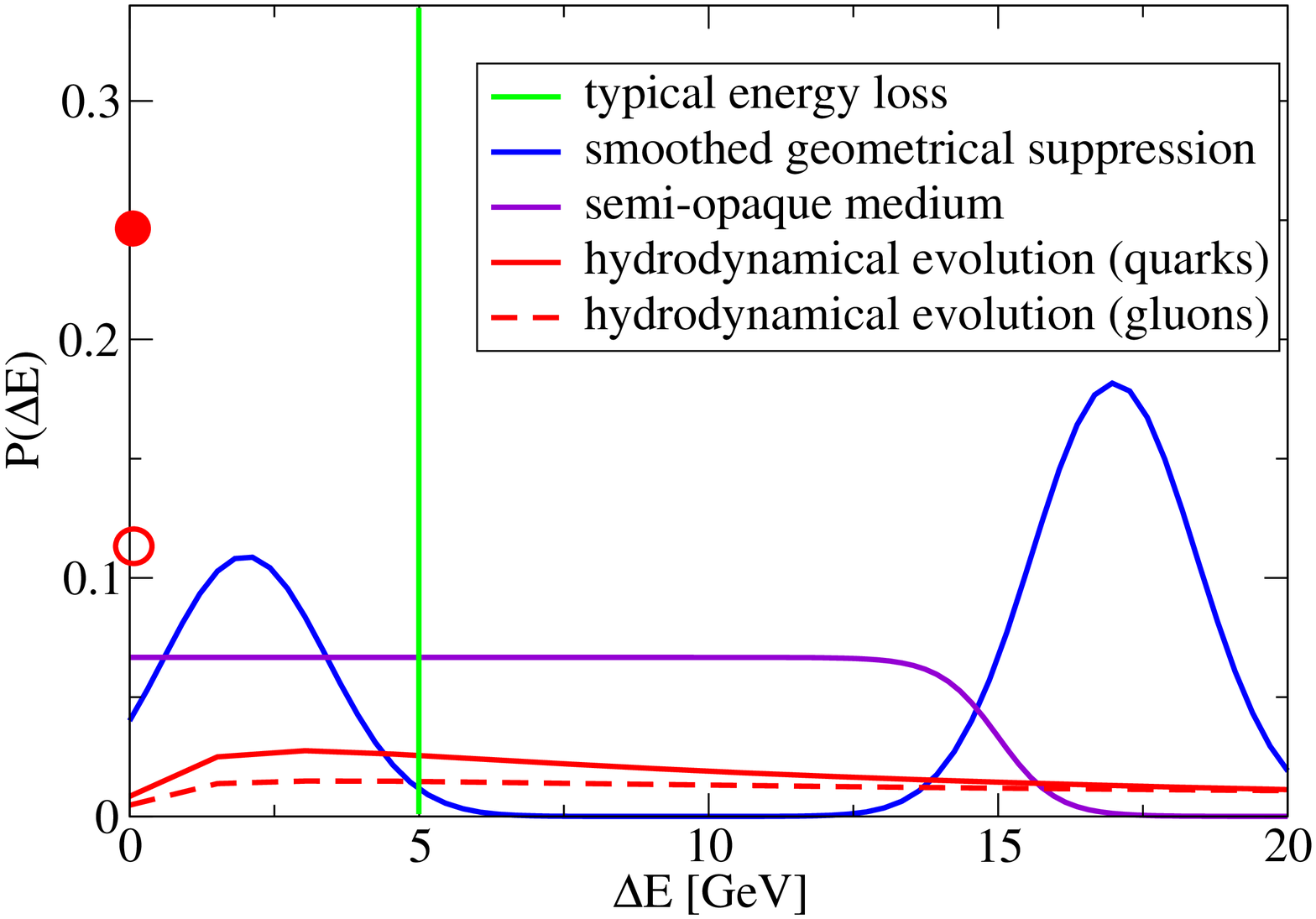, width=7.5cm}\epsfig{file=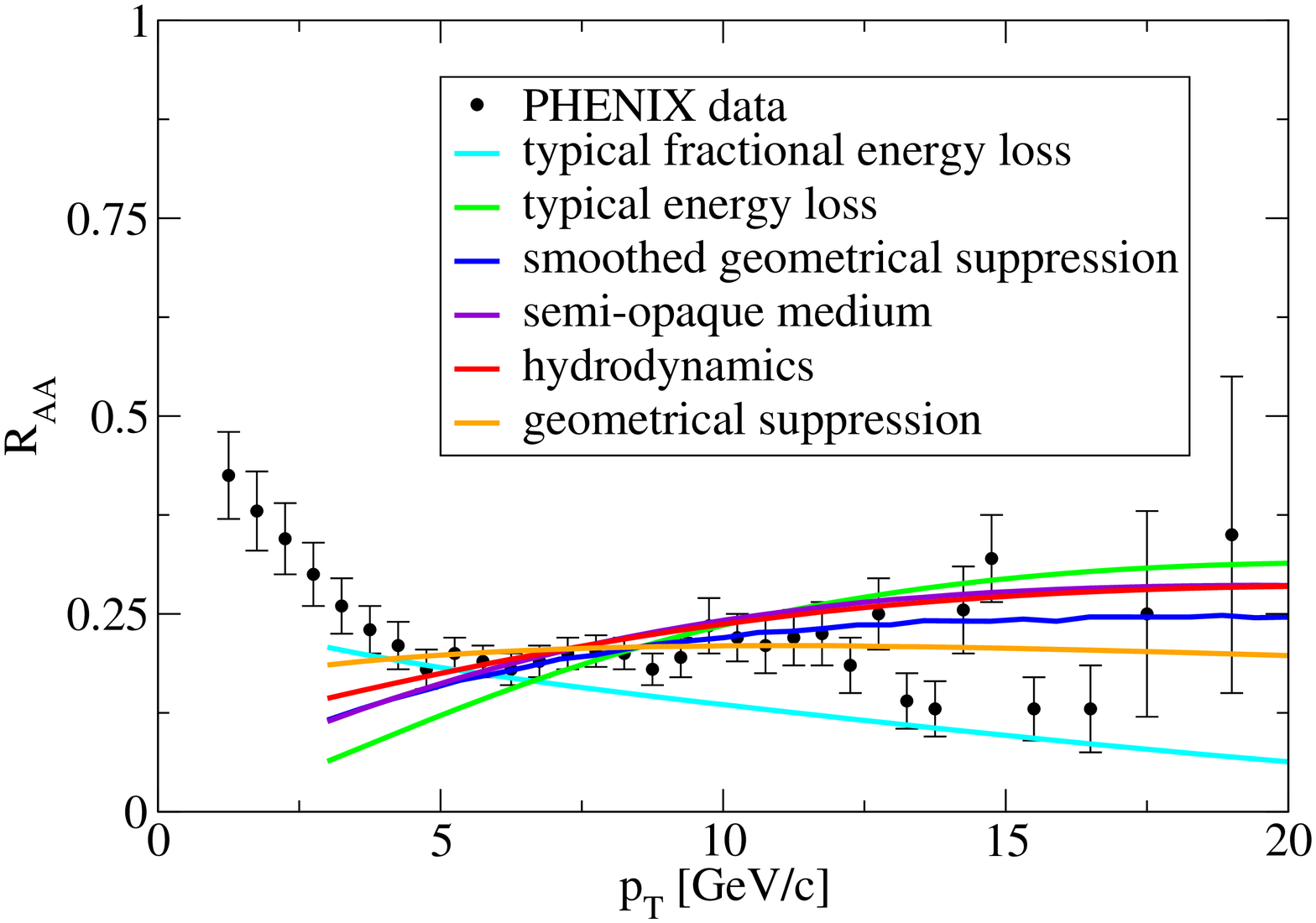, width=7.5cm}
 \caption{\label{F-1}Left panel: Trial energy loss distributions $\langle P(\Delta E) \rangle_{T_{AA}}$ for various scenarios of jet energy loss in the medium (see text and \cite{gamma-h} for details). Right planel: $R_{AA}$ as calculated from the trial distributions shown on the left hand side.}
\end{figure}

Before we proceed to calculate this quantity, let us illustrate the sensitivity of $R_{AA}$ to details of $\langle P_f(\Delta E)\rangle_{T_{AA}}$ (and hence the potential for tomographic information) by inserting trial distributions into the folding integral Eq.~(\ref{E-Folding}). These trial distributions are shown in Fig.~\ref{F-1}, left panel, the resulting $R_{AA}$ is shown in the right panel and compared with the PHENIX data for pions \cite{PHENIX_R_AA} (see also \cite{gamma-h} for details).

It is apparent from the figure that despite strong differences in the functional form of $\langle P(\Delta E) \rangle_{T_{AA}}$, all distributions describe the measured $R_{AA}$ reasonably well above some minimum $p_T$. The notable exception is the case of a constant fractional energy loss in which $R_{AA}$ drops as a function of $p_T$, which does not seem to capture the overall trend well. It has to be concluded that $R_{AA}$ does not exhibit great tomographic capability beyond a single overall energy loss scale (the numerical value of which moreover is different for each model). This may explain why different calculations extract rather different quenching properties of the medium from fits to $R_{AA}$. 

However, while the curves are reasonably similar over the kinematic range shown here, they do show differences in details which unfortunately cannot be resolved within the current data precision. Thus, there is some reason to suspect that either increased ecperimental statistics or a larger accessible kinematic range may provide more stringent constraints for the energy loss distribution.

\begin{figure}
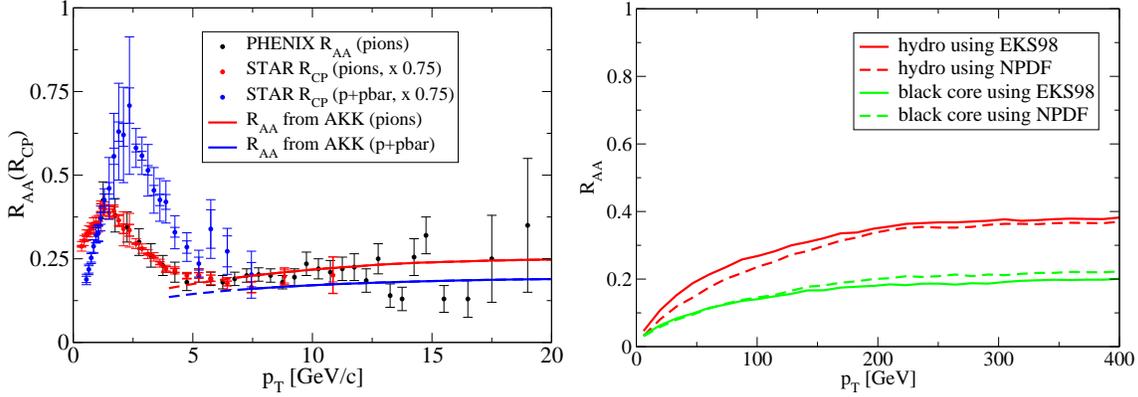

\epsfig{file=R_AA_protons.eps, width=7.4cm}\hspace*{0.2cm}\epsfig{file=R_AA_LHC.eps, width=7.4cm}
 \caption{\label{F-2}Left panel: $R_{AA}$ for pions and protons as compared to the measured $R_{AA}$ ($R_{CP}$) data \cite{PHENIX_R_AA,STAR_R_AA} for RHIC conditions  Right planel: Model predictions for $R_{AA}$ at the LHC based on two scenarios which describe the data at RHIC \cite{LHC-prediction}. }
\end{figure}

In Fig.~\ref{F-2} left panel we show the calculated $R_{AA}$ using the procedure outlined above to determine $\langle P(\Delta E) \rangle_{T_{AA}}$ instead of a trial ansatz. Once $K$ is adjusted, the result does not exhibit strong sensitivity to the underlying medium evolution model (we do not show the results for all different medium evolutions here as the curves are difficult to distinguish), again confirming that $R_{AA}$ has very limited tomographic capability in the RHIC kinematic range.

If the AKK fragmentations \cite{AKK} are used for computation for which the baseline process of proton production in p-p collisions is roughly under control (in detail, AKK seems to overpredict the process by about a factor 2), the calculation of  $R_{AA}$ for both pions and protons agrees well with the data \cite{R_AA_ppbar}. This is not a trivial result, as in the AKK fragmentation scenario proton production is gluon-dominated whereas pion production is not, hence the difference between pion and proton production should reflect the different energy loss properties of quarks and gluons. In the calculation as presented here, the rather small difference between proton and pion suppression is caused by the fact that gluon suppression is already in a saturated regime --- increasing the quenching power of the medium further induces only a small change in the gluonic $R_{AA}$ \cite{R_AA_ppbar}.

In Fig.~\ref{F-2}, right panel we show the extrapolation of the $p_T$ dependence of $R_{AA}$ to LHC energies based on the hydrodynamical scenarios which describe the data at RHIC \cite{Hydro,LHC-prediction}. While there is some uncertainty associated with the extrapolation of the nuclear parton distribution function (NPDF \cite{NPDF} vs. EKS98 \cite{EKS98}), this is a small effect, and it becomes indeed apparent that with the extended kinematic lever-arm of LHC the different properties of the two scenarios (dense core and dilute halo vs. more evenly distributed quenching power) can clearly be distinguished. The results here differ from a previous calculation presented in \cite{Fragility}. The improvement of the present calculation over the previous work is chiefly in the use of a dynamically evolving soft medium instead of a static cylinder ansatz and in using Eq.~(\ref{E-PGeo}) for the primary vertex distribution as compared to a homogeneous distribution.

\begin{figure}
\epsfig{file=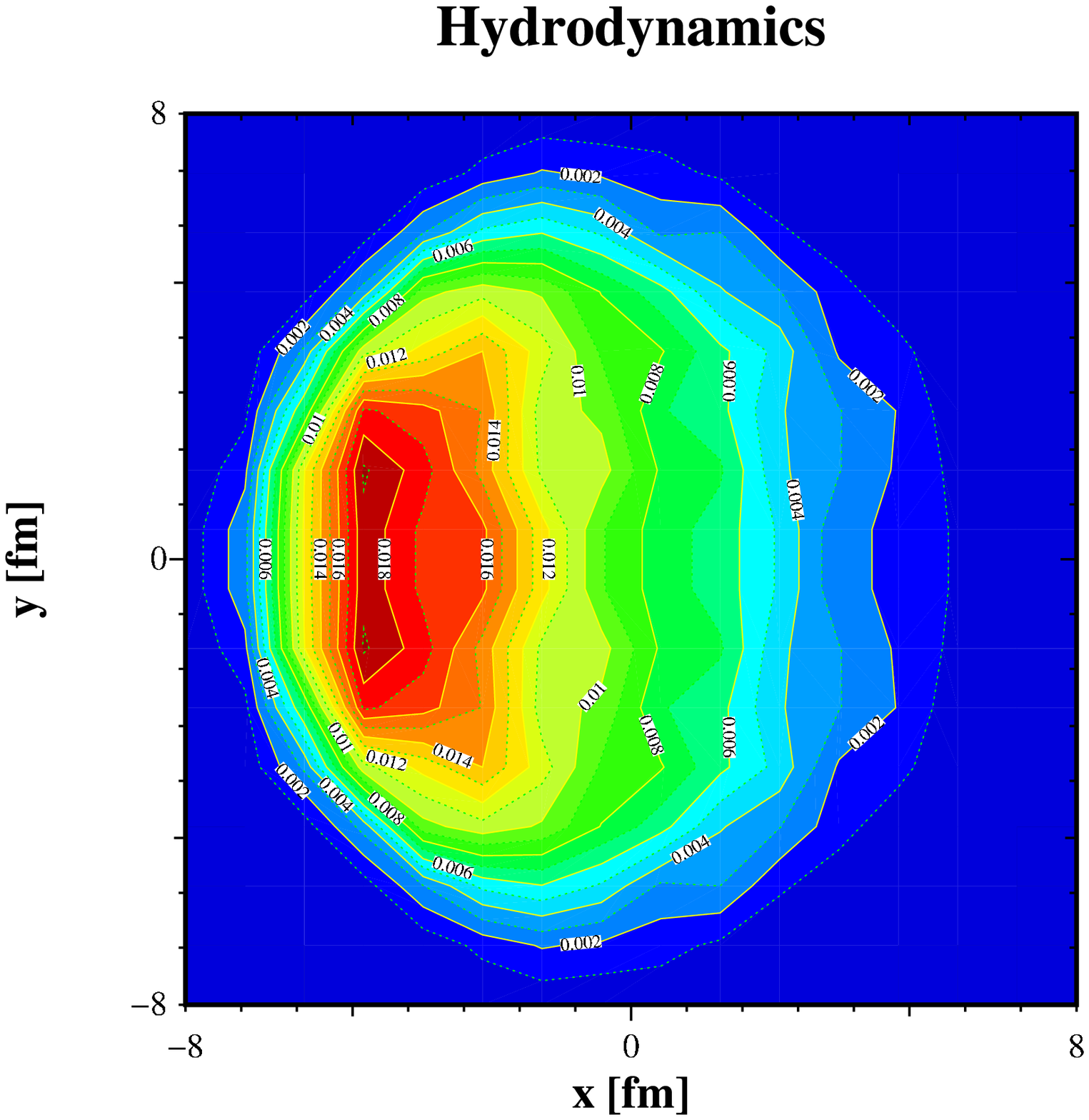, width=5.3cm}\hspace{-0.3cm}\epsfig{file=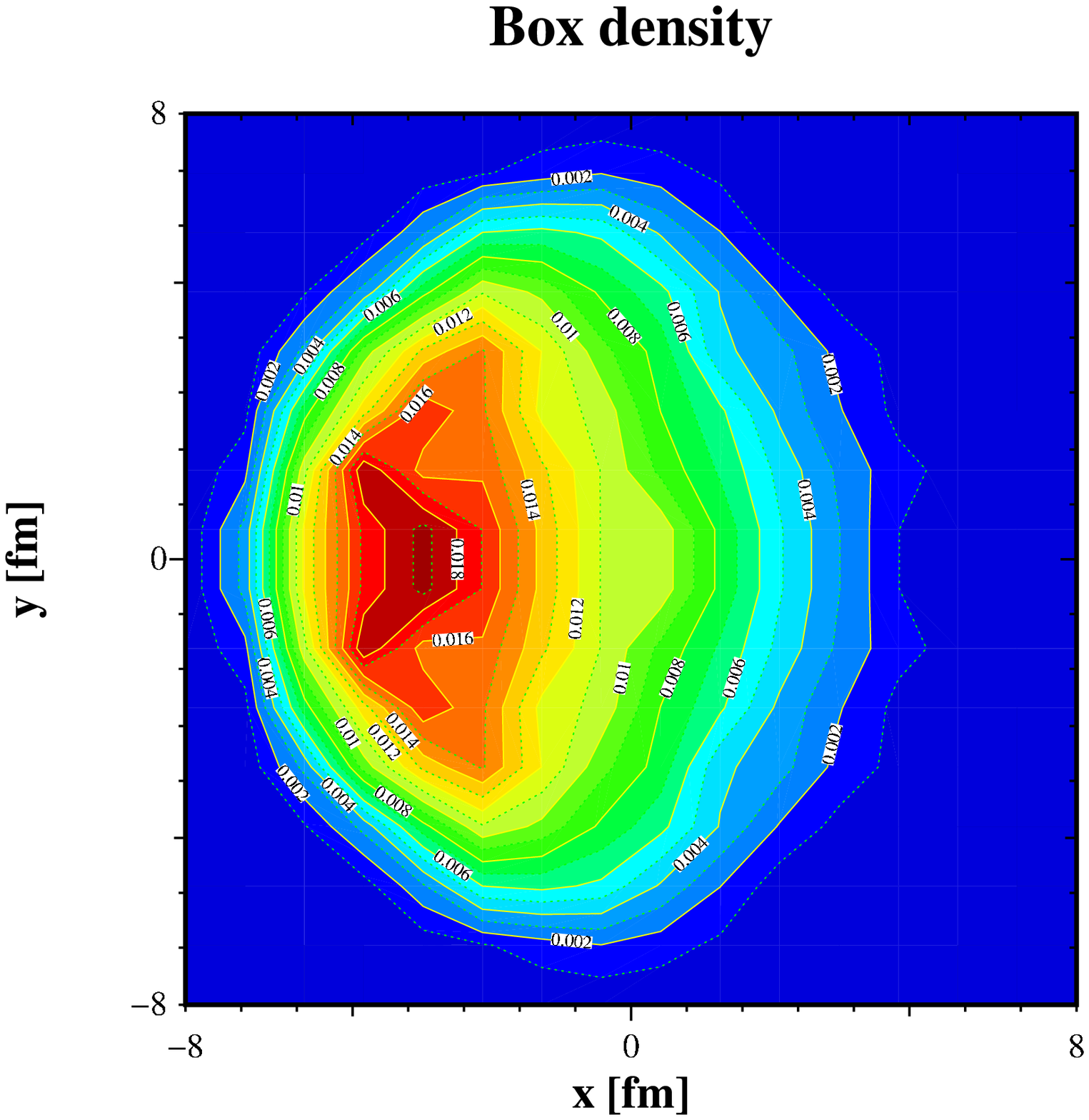, width=5.3cm}\hspace{-0.09cm}\raisebox{1.5cm}{\epsfig{file=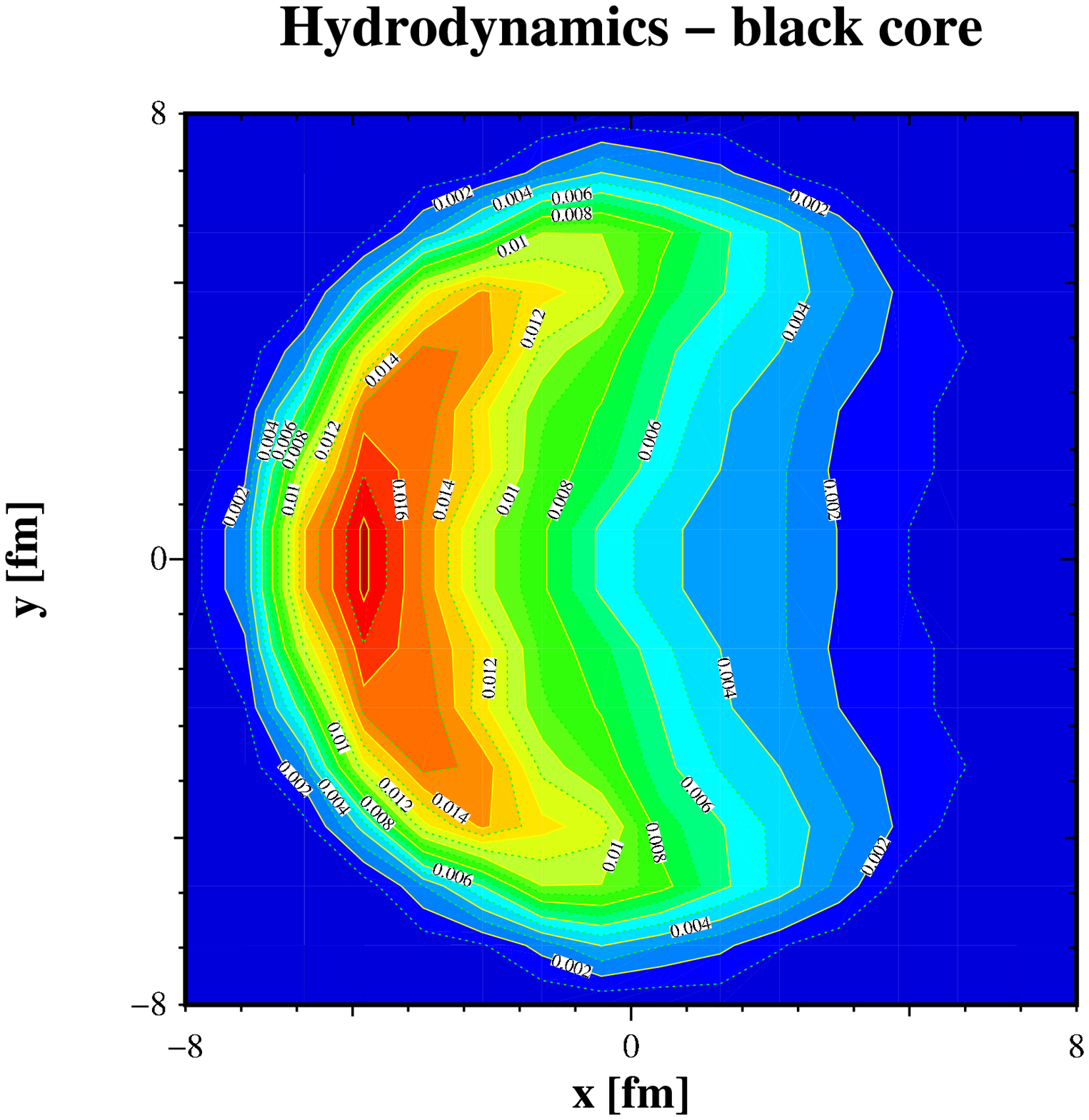, width=4.91cm}}
\vspace*{-2cm}
\caption{\label{F-3}Probability density for finding a hard vertex in the transverse plane in 200 AGeV Au-Au collisions leading to an observed hadron above 8 GeV transverse momentum, shown for three different medium evolution scenarios (see text). In all cases, the hard hadron propagation defines the $-x$ direction. All contour intervals are linear.}
\end{figure}

 Let us illustrate the differences induced by the spatial distribution of the quenching power by studying the geometry of single hadron suppression directly in the model.
In Fig.~\ref{F-3} we show the probability density of finding the primary pQCD vertex leading to an observed hadron above 8 GeV in $p_T$. It is evident (and quite expected) that emission occurs predominantly close to the near side surface of the medium. However, the degree to which surface emission is realized is quite different in all three models. Clearly, the strong suppression from the core region of the black core scenario repels the distribution much more from the center than the more even distribution of the other two scenarios. Thus, surface emission is not a property of a particular energy loss formalism, but arises from the interplay of energy loss formalism with the underlying geometry and evolution of the soft medium.

\section{Dihadron suppression}

We can make use of the sensitivity of the vertex distribution of single hadron suppression to the medium evolution by considering dihadron suppression. In a back-to-back event, the second hadron propagation path is not averaged over the initial overlap Eq.~(\ref{E-PGeo}) but over a conditional probability distribution {\it} given a valid trigger, i.e. over the distribution shown in Fig.~\ref{F-3} (which is quite different from the overlap). Thus, even if two model evolutions lead to identical $R_{AA}$, this does not mean that they would produce the same dihadron correlation pattern. We call this conditional probability distribution given a high-$p_T$ near-side trigger in the following $\langle P(\Delta E)\rangle_{Tr}$ and investigate its capability to obtain tomographical information.

For computational purposes, we employ a Monte-Carlo (MC) simulation of the experimental trigger condition, followed by the simulation of the away-side parton intrinsic-$k_T$ smearing, propagation, energy loss and fragmentation. The procedure is described in detail in \cite{Correlations}.

\begin{figure}
\epsfig{file=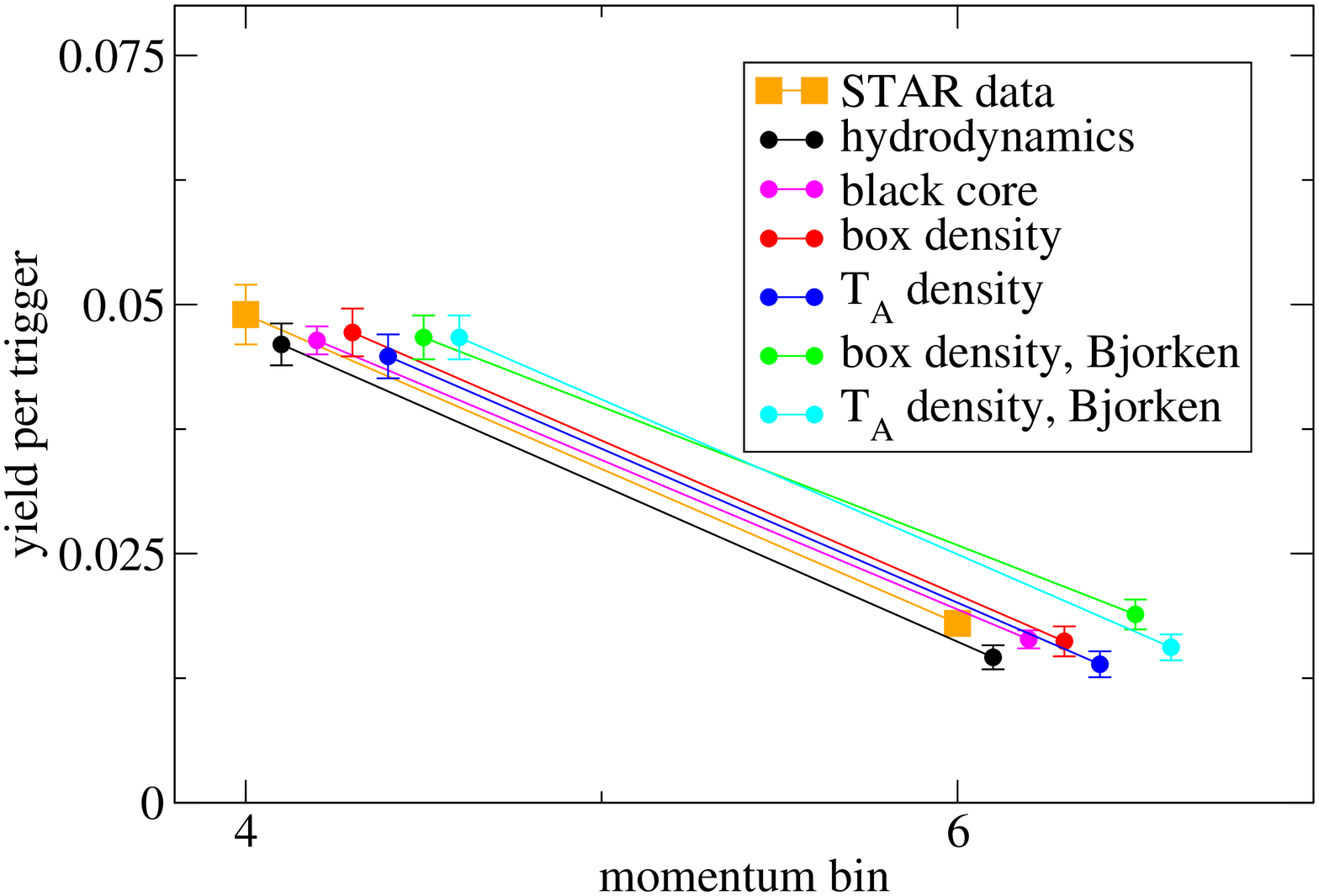, width=7.5cm}\epsfig{file=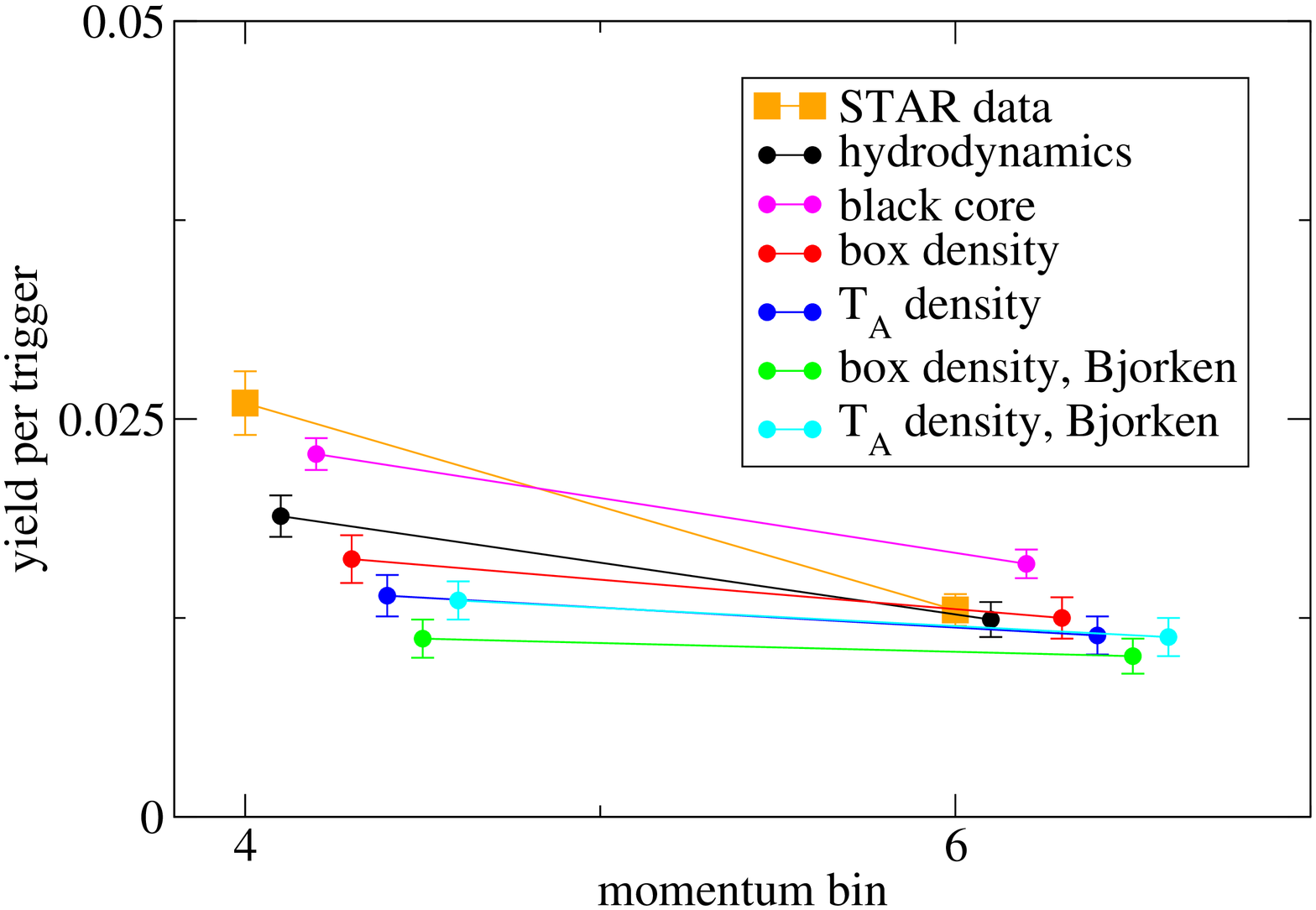, width=7.5cm}
\caption{\label{F-4}Yield per trigger on the near side (left panel) and away side (right panel)  
of hadrons in the 4-6 GeV and 6+ GeV momentum bin associated with a trigger in the range 8 GeV $<  
p_T < $ 15 GeV for the different models of spacetime evolution as compared with the STAR data  
\cite{Dijets1,Dijets2}. The individual data points have been spread artificially along the $x$ axis  
for clarity of presentation.}
\end{figure}

In Fig.~\ref{F-4} we compare the yield per trigger on the near and away side for different medium models with the data obtained by the STAR collaboration \cite{Dijets1,Dijets2}.
Within errors, the near side yield per trigger is described by all the models well. There is no 
significant disagreement among the models. The model calculations appear significantly more different if we consider the away side yield. 
Here, results for the 4-6 GeV momentum bin differ by almost a factor two.  However, none of the 
model calculations describes the data in this bin. This is in fact not at all surprising as below 
~5 GeV the inclusive single hadron transverse momentum spectra are not dominated by pQCD 
fragmentation and energy losses but, rather, by hydrodynamics possibly supplemented with 
recombination \cite{Reco,Coalescence} type phenomena. 
For this reason, the ratio $R_{AA}$ at $p_T< 5$~GeV cannot be expected to be described by pQCD 
fragmentation and energy losses, either. 

This is clearly unfortunate, as the model results are considerable closer to the experimental 
result in the 6+ momentum bin on the away side and hence our ability to discriminate between different models
is reduced. Since at this large transverse momenta the pQCD 
fragmentation + energy losses dominate the singe hadron spectrum, we expect that the model is able to give a valid description of the relevant physics 
in this bin: Not only is $R_{AA}$ well described by the data, but also the contribution of 
recombination processes to the yield is expected to be small \cite{Reco}. Thus, as it stands, only 
the black core scenario can be ruled out by the data, the box density with Bjorken expansion seems 
strongly disfavoured but still marginally acceptable.

Thus, as it stands, the kinematic window to study dihadron correlations in a perturbatively calculable region is not enough to exploit the difference between $\langle P(\Delta E)\rangle_{Tr}$ and $\langle P(\Delta E)\rangle_{R_{AA}}$ and thus to obtain  detailed tomographic information. However, the situation may improve for an increased kinematical window in the region where pQCD + fragmentation can be applied. In order to test this, we redo the MC simulation with trigger hadrons in the range between 12 and 20 GeV.

\begin{figure}
\epsfig{file=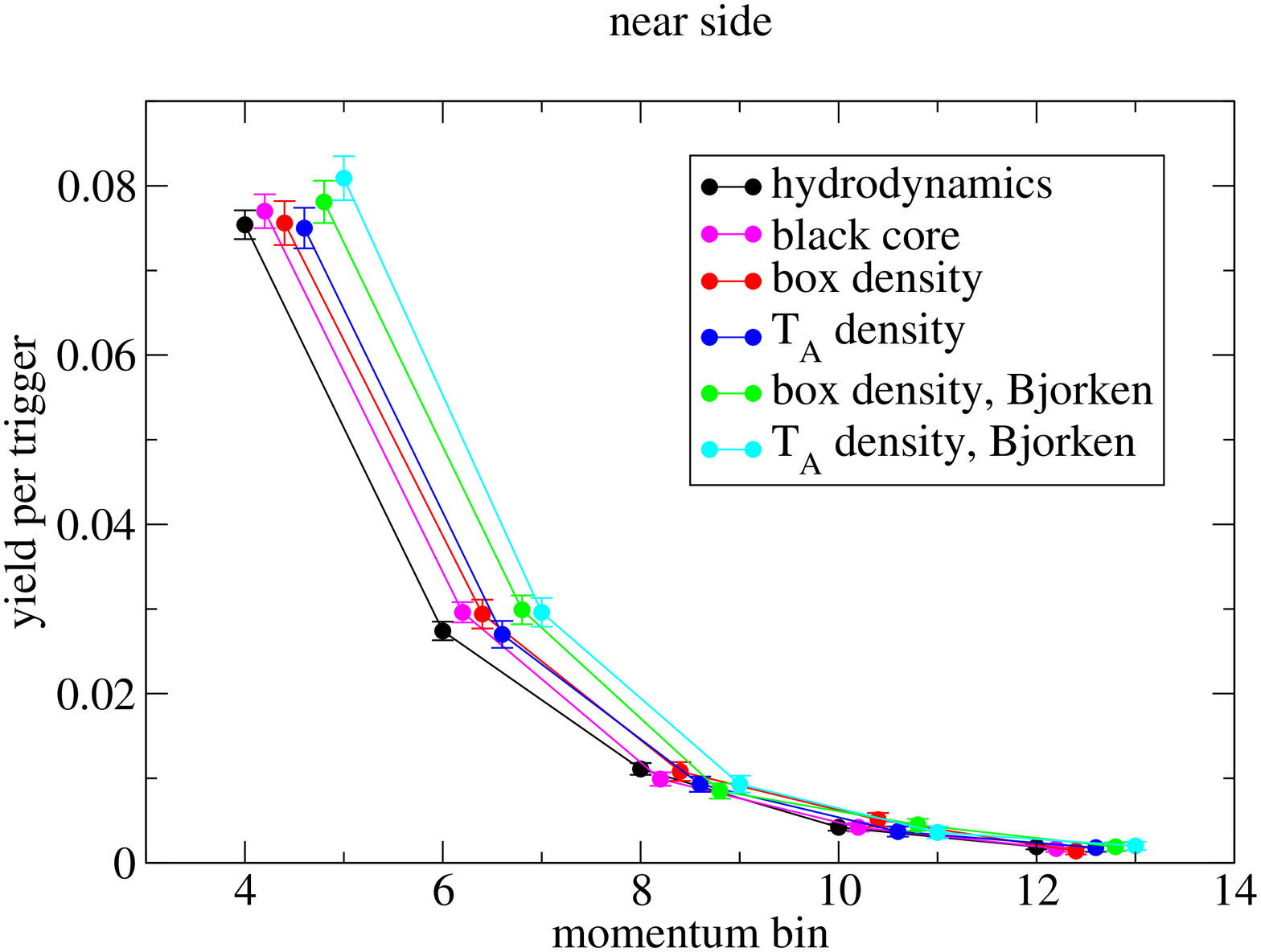, width=7.5cm}\epsfig{file=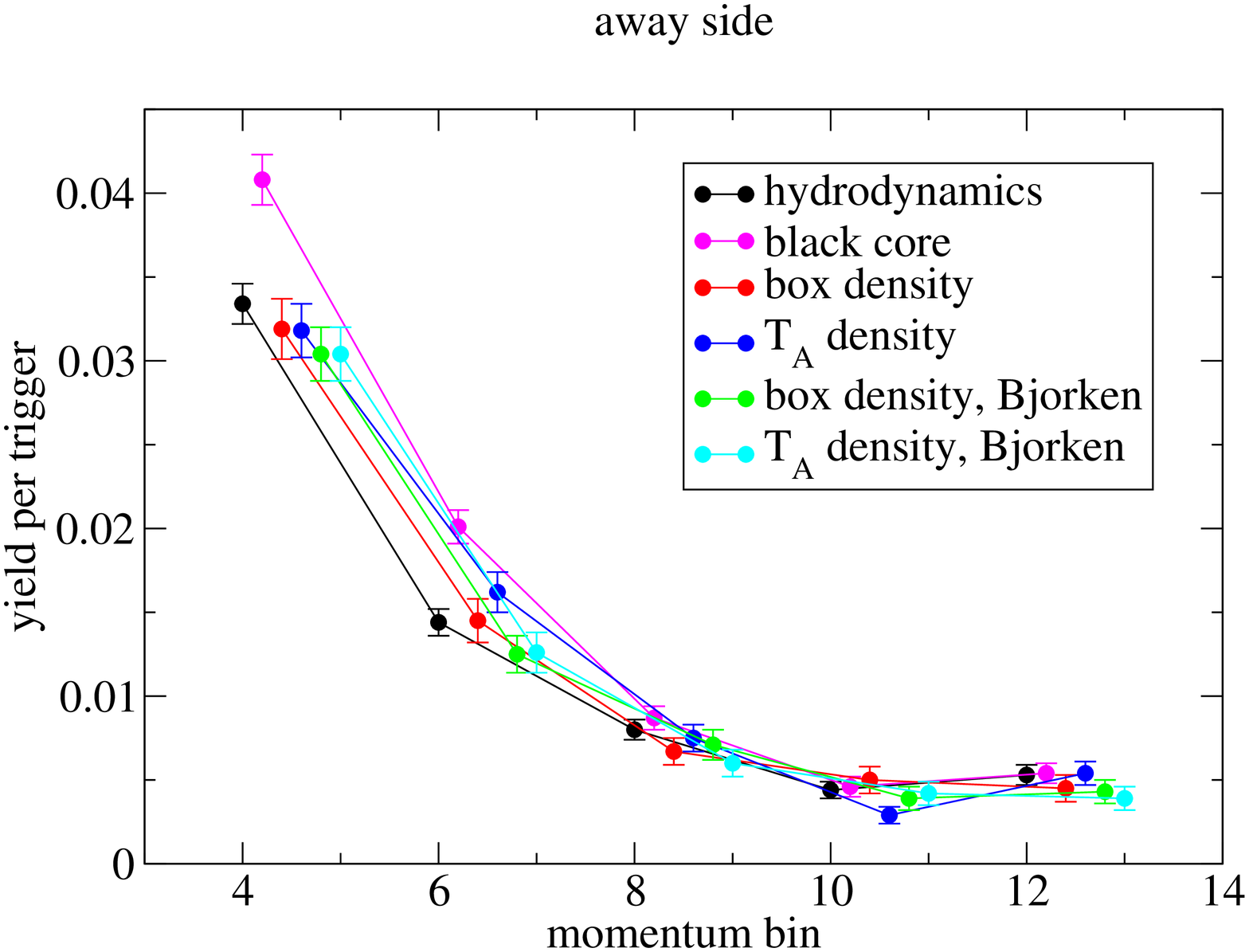, width=7.5cm}
\caption{\label{F-5}Yield per trigger on the near side (left panel) and away side (right panel)  
of hadrons in the 4-6 GeV and 6+ GeV momentum bin associated with a trigger in the range 12 GeV $<  
p_T < $ 20 GeV for the different models of spacetime evolution. The individual data points have been spread artificially along the $x$ axis  
for clarity of presentation.}
\end{figure}

The distribution after fragmentation into hadrons in bins of 2 GeV width in the perturbative region  
is shown in Fig.~\ref{F-5} for the near side (left panel) and away side (right panel). It is  
again apparent that within errors all models agree in the expected near side yield. The momentum  
spectrum of the away side exhibits considerably more structure. Several of the scenarios can now be  
clearly told apart in bins in the perturbative region. For example the $T_A$ and the box density  
(which have virtually identical $\langle P(\Delta E)\rangle_{T_{AA}}$)  show  
almost a factor two difference in the 10-12 GeV momentum bin.
As we have seen above in the case of the LHC extrapolation, is evident again from the analysis that having a larger lever-arm in momentum is needed to get access to tomographic information.

\begin{figure}
\epsfig{file=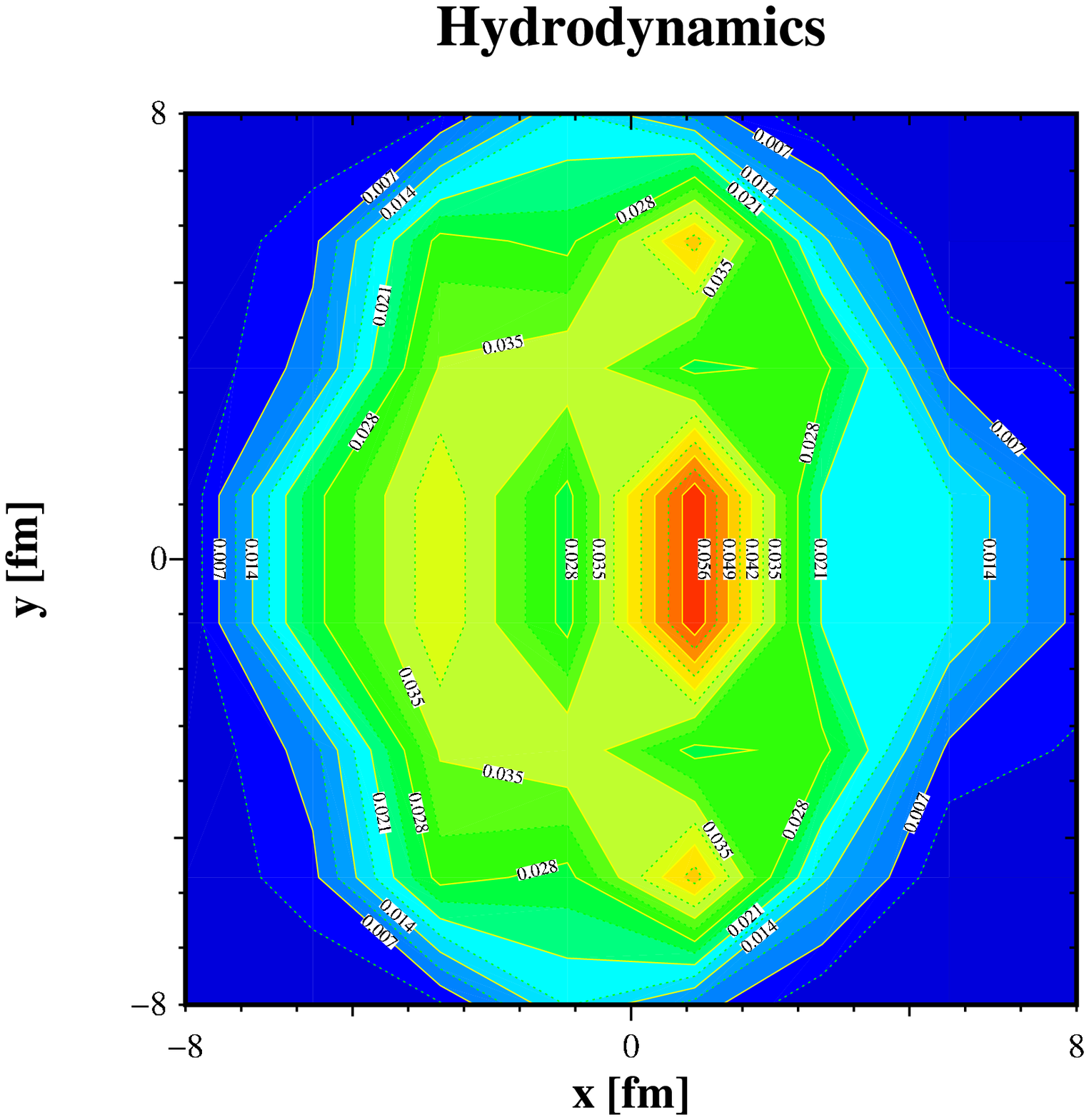, width=5.2cm}\hspace{-0.3cm}\epsfig{file=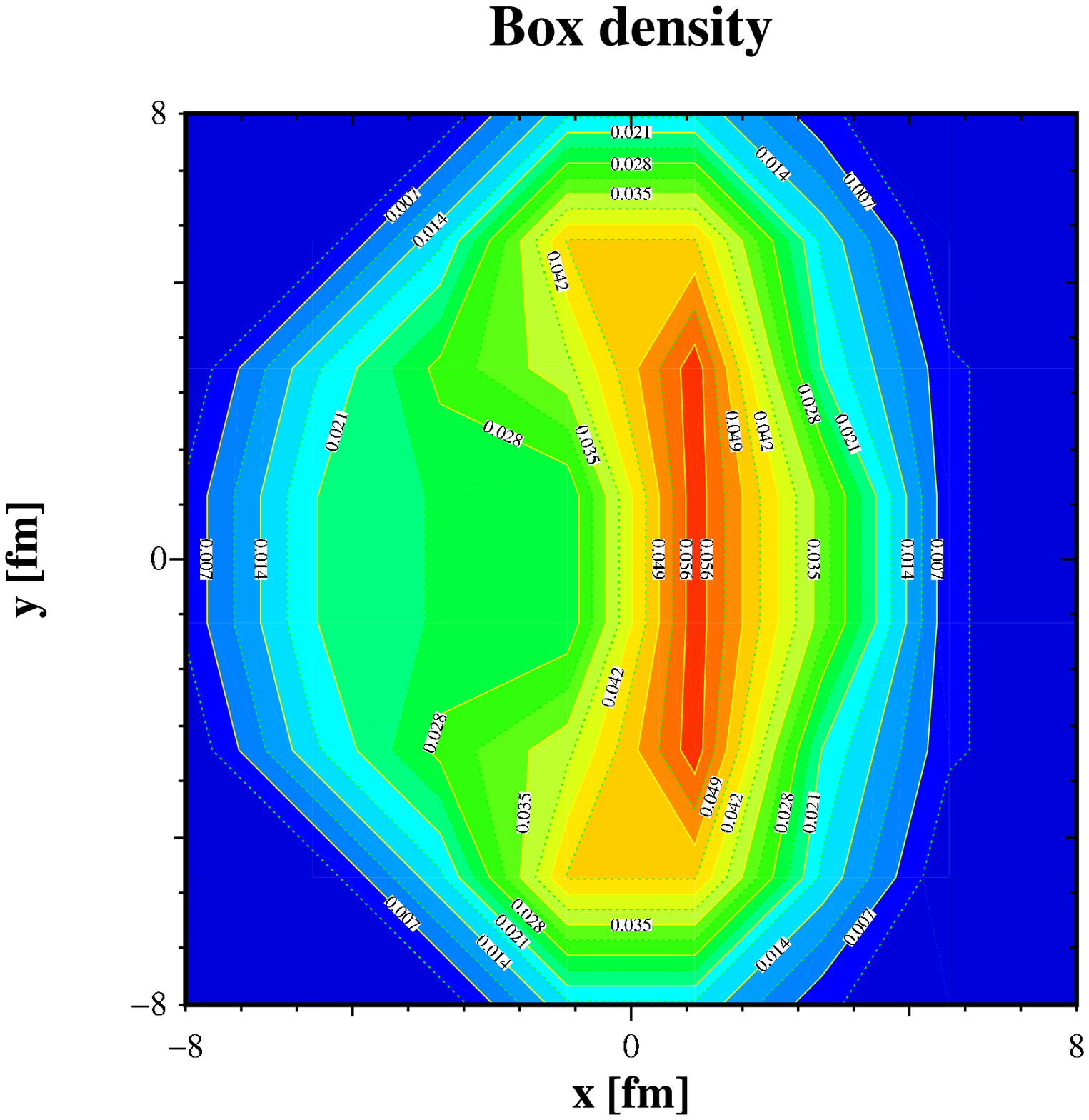, width=5.2cm}\hspace{-0.3cm}\epsfig{file=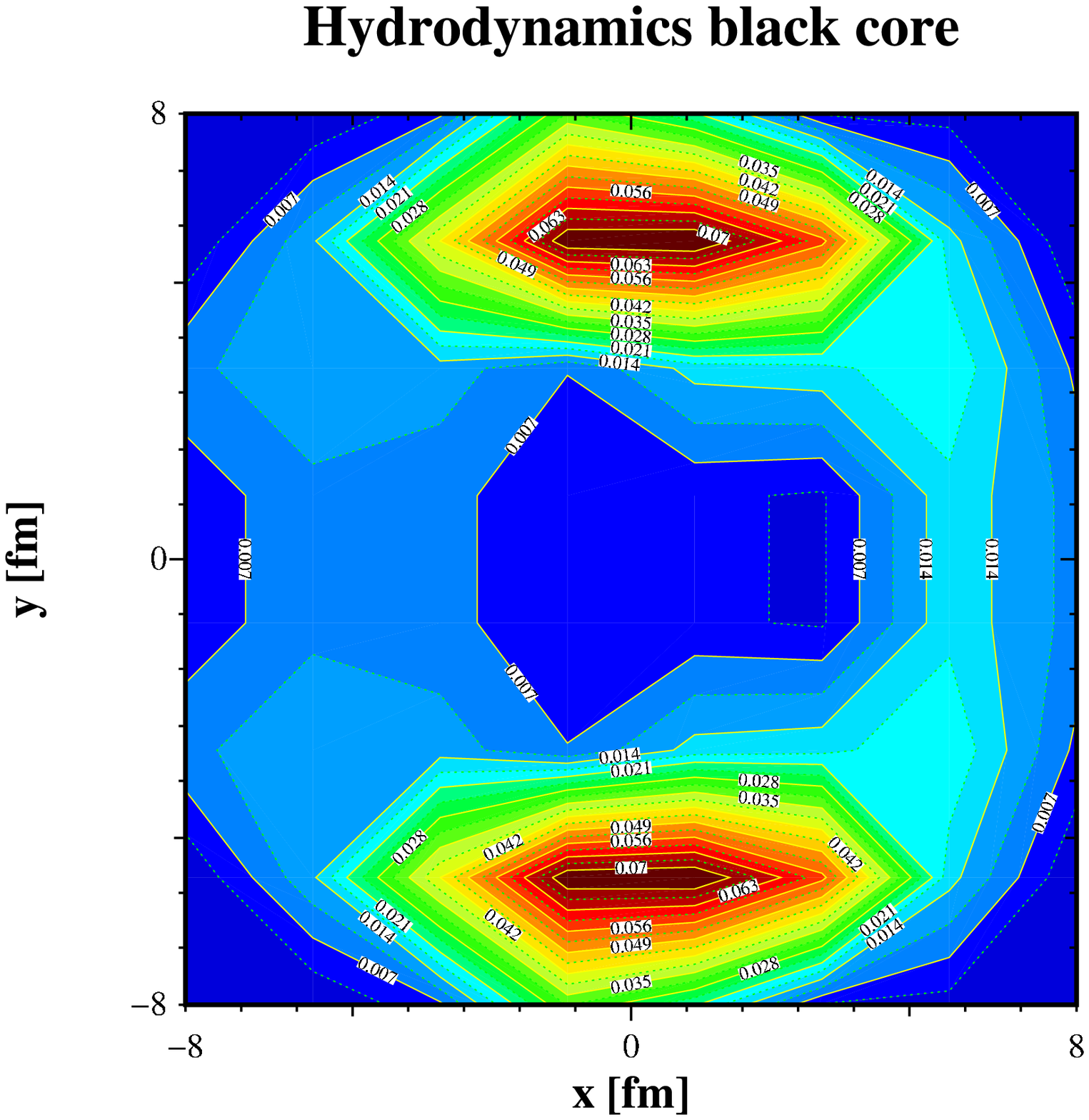, width=5.2cm}
\vspace*{-1.8cm}
\caption{\label{F-6}Probability density for finding a hard vertex in the transverse plane in 200 AGeV Au-Au collisions leading to an event with both an observed near side hadron above $p_T =8$ GeV (defining the $-x$ direction) and an away side hadron with $p_T > 4$ GeV for three different medium evolutions (see text). All contour intervals are linear.}
\end{figure}

Finally, let us discuss the geometry of dihadron suppression. In Fig.~\ref{F-6} we show the probability density of finding a hard vertex leading to a high $p_T$ near side trigger and a correlated associated hard away side hadron. Here, clear differences between tangential emission in the case of a dense core and production across the whole volume become apparent.

\begin{figure}
\begin{center}
\epsfig{file=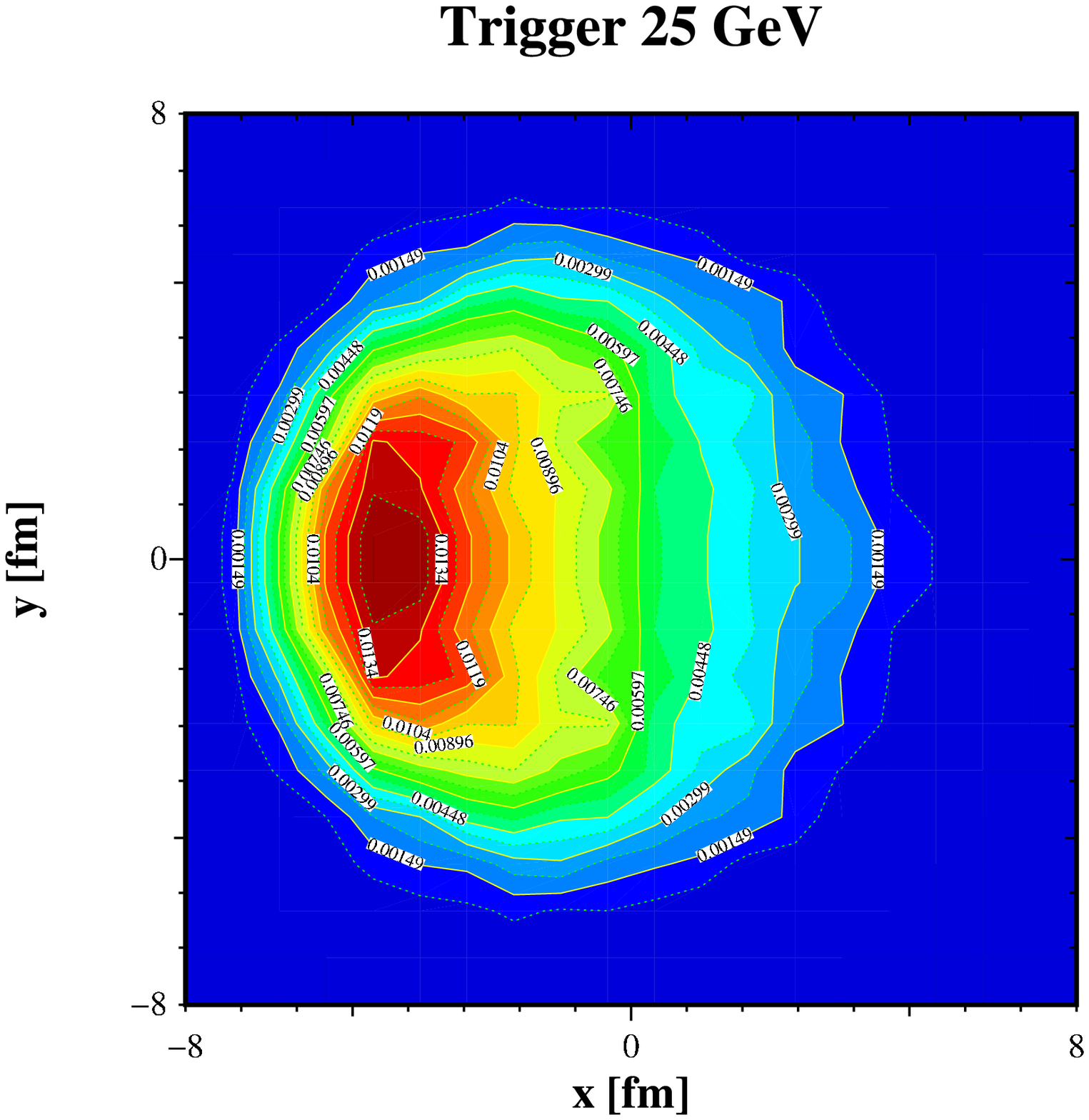, width=6cm}\epsfig{file=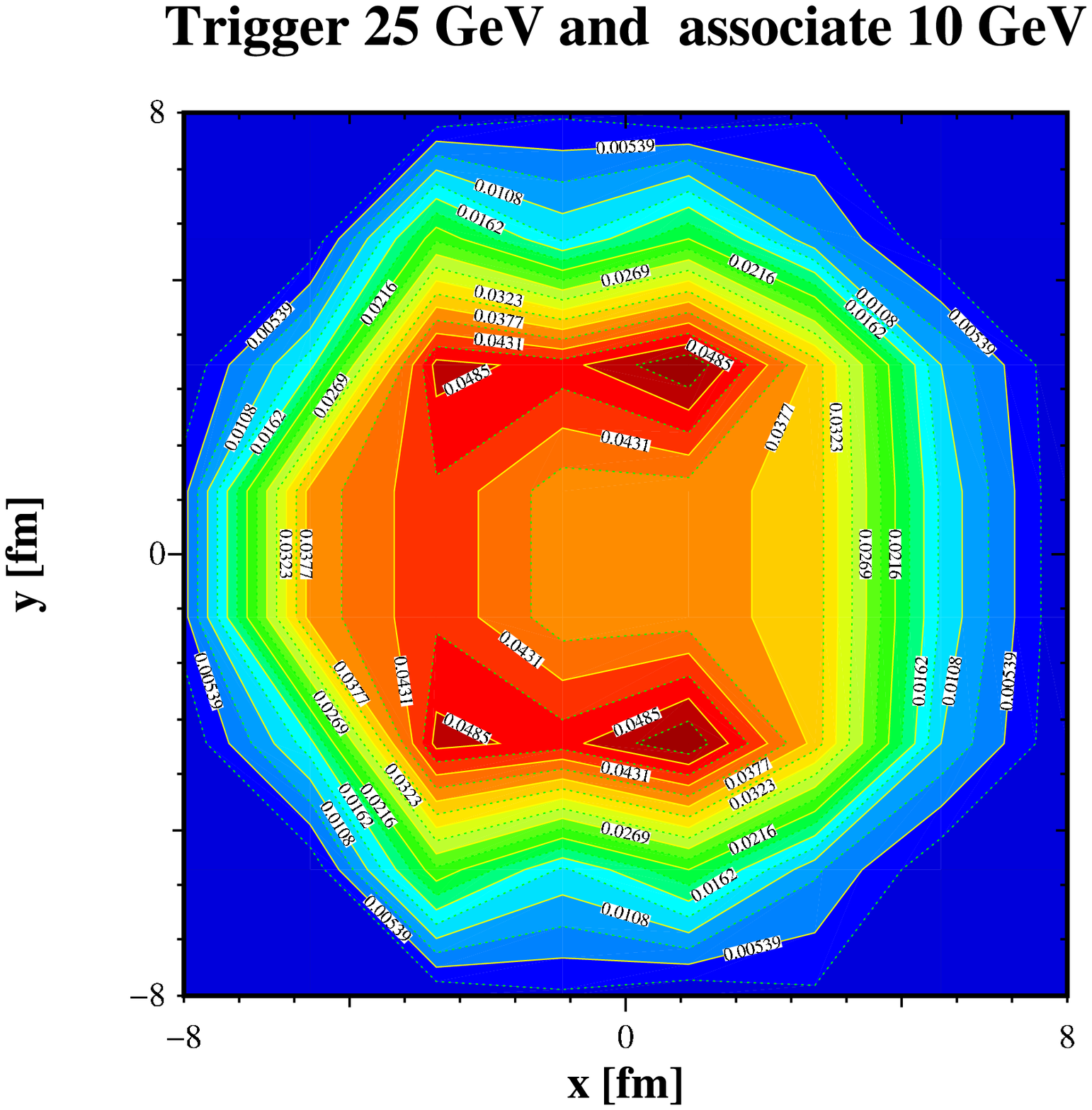, width=6cm}
\end{center}
\vspace*{-2cm}
\caption{\label{F-7}Left panel: Probability density for finding a hard vertex in the transverse plane in 5.5 ATeV Pb-Pb collisions leading to a near side hadron with $p_T > 25$ GeV propagating into the $-x$ direction for the LHC hydrodynamical model prediction. Right: Probability density requiring in addition an associated away side hadron with $p_T > 10$ GeV momentum. All contour intervals are linear. }
\end{figure}

In Fig.~\ref{F-7} we also show the geometry of single hadron and dihadron suppression for LHC conditions in central Pb-Pb collisions. For a 25 GeV trigger hadron, we expect some degree of surface emission (note that the dihadron production distribution is somewhat repelled from the center) but no strong tangential emission.

\section{A simple model}

As we have seen, quite a general class of models predict a small rise of $R_{AA}$ with $p_T$ at RHIC and a more pronouned one at LHC. Assuming RHIC kinematics, $R_{AA}$ is rather insensitive to details of the energy loss probability distributions, at LHC the sensitivity is considerably enhanced. Likewise, dihadron correlations become more sensitive to the medium density distribution if the kinematic range is increased. In the following, let us try to illustrate that all these observations can be understood from simple considerations.

Quite generally, energy loss probability distributions can be decomposed as
\begin{equation}
\langle P(\Delta E) \rangle_{T_{AA}(Tr)} = T \delta(\Delta E) + S \cdot P(\Delta E) + A \cdot \delta(\Delta E - E)
\end{equation}
where $T$ is a transmission term describing a parton penetrating through the medium without energy loss, $S$ is a shift term which characterizes partons emerging from the medium after a finite energy loss and $A$ is an absorption term describing partons which have been shifted in energy so much that they become part of the soft medium.

Let us now assume a power law for the parton spectrum at RHIC and LHC $\sim 1/p_T^n$ with $n_{RHIC} > n_{LHC}$. Energy loss $\Delta E$ then changes this spectrum to  $1/(p_T + \Delta E)^n$, thus $R_{AA}$ in this simple model can be obtained from 
\begin{displaymath}
R_{AA} \approx \int d\Delta E \langle P(\Delta E) \rangle_{T_{AA}} 1/\left(1+\frac{\Delta E}{p_T}\right)^n
\end{displaymath}

It is evident from the expression that $R_{AA}$ at given $p_T$ is equal to the transmission term $T$ plus a contribution which is proportional to the integral of $\langle P(\Delta E) \rangle_{T_{AA}}$ from zero up to the energy scale $E_{max}$ of the parton, {\it seen through the filter} of the steeply falling spectrum. $R_{AA}$ grows with $p_T$ since $E_{max}$ grows linearly with $p_T$. However, at RHIC conditions the characteristic scale $\omega_c$ of the energy loss probability distribution is far above $E_{max}$, thus the growth is slow and $R_{AA}$ is dominated by $T$, rendering it almost a constant. Since tomographic information is mainly contained in the shift term $S$, the apparent insensitivity of $R_{AA}$ to assumptions about the medium can be understood.

This is very different at LHC where $E_{max} \sim \omega_c$ (since $E_{max}$ grows linear with $p_T$ but $\omega_c$ grows with the entropy density and hence much slower) and a pronounced contribution of the shift term can be probed. Here, a rise of $R_{AA}$ with $p_T$ is expected, along with a greater tomographic sensitivity.

\section{Conclusions}

We have investigated the capability of single and dihadron suppression to provide tomographic information about the soft medium created in ultrarelativistic heavy-ion collisions. We have argued that at RHIC kinematics, the nuclear suppresssion factor $R_{AA}$ is not very sensitive to the medium evolution. While dihadron suppression, due to its different geometrical averaging, exhibits in principle more sensitivity to medium properties, unfortunately the present data situation allows only to rule out a very pronounced difference between a strongly absorbing core and a dilute halo. This insensitivity can be traced back to the fact that $\omega_c$, the intrinsic scale for energy loss is much higher than $E_{max}$, the accessible parton energy at RHIC.

However, when going to LHC energies, this condition no longer holds. $R_{AA}$ becomes dominated by partons being shifted in energy, and tomographic information can be recovered even from the $p_T$ dependence of $R_{AA}$. Dihadron correlations and other measurements, such as $\gamma$-hadron correlations \cite{gamma-h} which provide a monochromatic source of hard quarks in the medium or $R_{AA}$ vs. reaction plane \cite{R_AA_RP}, which allows for a systematic variation of in-medium pathlength, may provide additional information such that a multi-pronged approach to jet tomography finally becomes feasible.

\end{document}